\documentclass[
reprint,
superscriptaddress,
nofootinbib,
amsmath,amssymb,
aps,
pra
]{revtex4-2}

\usepackage{lmodern}
\usepackage{graphicx}
\usepackage{dcolumn}
\usepackage{bm}
\usepackage{physics}
\usepackage{newtxtext,newtxmath}
\usepackage[up]{subfigure}
\usepackage{color}
\usepackage[usenames,dvipsnames]{xcolor}
\usepackage[breaklinks, pdftex, hyperfootnotes=true, pdfpagelabels, bookmarks, pageanchor]{hyperref}
\usepackage{ulem}
\renewcommand{\doi}[2]{\href{http://doi.org/#1}{#2}}
\newcommand{\arxiv}[1]{\href{https://arxiv.org/abs/#1}{arXiv:#1}}
\hypersetup{
	colorlinks=true, linktocpage=true, pdfstartpage=1, pdfstartview=FitH, pdfborder={0 0 0},
	breaklinks=true, pdfpagemode=UseNone, pageanchor=true, pdfpagemode=UseOutlines,
	plainpages=false, bookmarksnumbered, bookmarksopen=true, bookmarksopenlevel=1,
	hypertexnames=true, pdfhighlight=/O,
	urlcolor=blue, linkcolor=blue, citecolor=blue,
}

\graphicspath{{./figure/}}
\begin{document}

\title{Variational optimization of the quantum annealing schedule for the Lechner-Hauke-Zoller scheme}

\author{Yuki Susa}
\email{y-susa@nec.com}
\affiliation{System Platform Research Laboratories, NEC Corporation, Kawasaki, Kanagawa 211-8666, Japan}
\affiliation{NEC-AIST Quantum Technology Cooperative Research Laboratory, National Institute of Advanced Industrial Science and Technology (AIST), Tsukuba, Ibaraki 305-8568, Japan}
\author{Hidetoshi Nishimori}
\affiliation{Institute of Innovative Research, Tokyo Institute of Technology, Yokohama, Kanagawa 226-8503, Japan}
\affiliation{Graduate School of Information Sciences, Tohoku University, Sendai 980-8579, Japan}
\affiliation{RIKEN Interdisciplinary Theoretical and Mathematical Sciences (iTHEMS), Wako, Saitama 351-0198, Japan}

\date{\today}

\begin{abstract}
The annealing schedule is optimized for a parameter in the Lechner-Hauke-Zoller (LHZ) scheme for quantum annealing designed for the all-to-all-interacting Ising model representing generic combinatorial optimization problems.  We adapt the variational approach proposed by Matsuura {\it et al.} (arXiv:2003.09913) to the annealing schedule of a term representing a constraint for variables intrinsic to the LHZ scheme with the annealing schedule of other terms kept intact. Numerical results for a simple ferromagnetic model and the spin-glass problem show that nonmonotonic annealing schedules optimize the performance measured by the residual energy and the final ground-state fidelity.  This improvement does not accompany a notable increase in the instantaneous energy gap, which suggests the importance of a dynamical viewpoint in addition to static analyses in the study of practically relevant diabatic processes in quantum annealing. 
\end{abstract}

\maketitle

\begin{figure*}[t]
\centering
\includegraphics[width =0.8\textwidth]{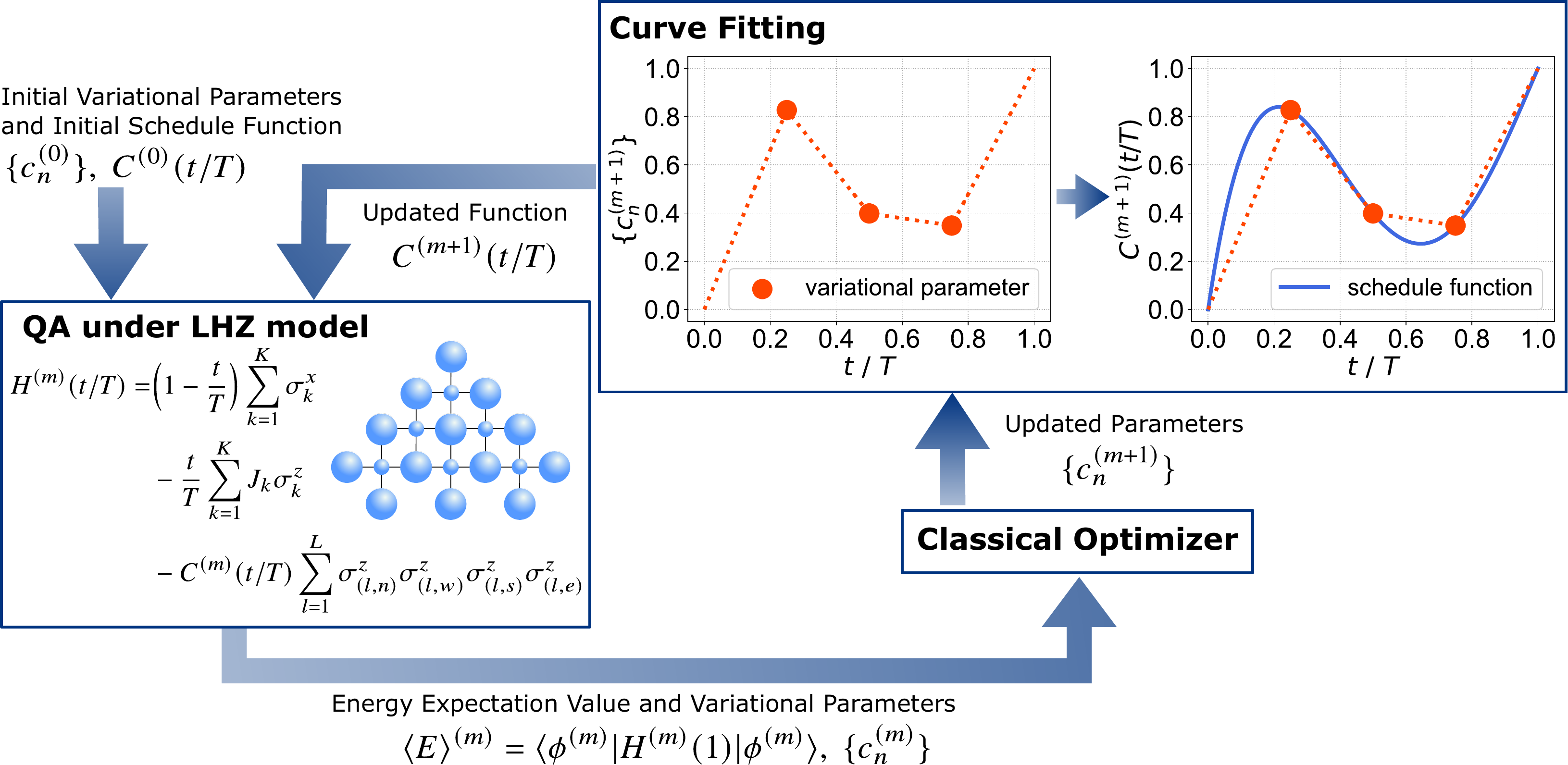}
\caption{Flowchart of the algorithm.}
\label{fig1}
\end{figure*}

\section{Introduction}
Quantum annealing (QA) is a quantum-mechanical metaheuristic for combinatorial optimization problems  \cite{kadowaki1998quantum,brooke1999quantum,santoro2002theory,santoro2006optimization,das2008colloquium,morita2008mathematical,hauke2020perspectives} and is related to adiabatic quantum computation \cite{farhi2001quantum,albash2018adiabatic}. An important goal of QA is to find the ground state of a classical Ising model, to which the cost function of a combinatorial optimization problem can be mapped \cite{lucas2014ising}.
One of the challenges in the experimental implementation of QA is the realization of  long-range interactions in the Ising model on locally connected qubits on the device, especially in the superconducting technology \cite{Harris2010}.
The standard approach to this problem is embedding \cite{choi2008minor,choi2011minor}, where a logical qubit is represented by a set of strongly coupled physical qubits, but it leads to an overhead in the number of qubits as well as possible errors caused by misalignment among physical qubits. Although the former issue still remains, the latter problem is mitigated by the scheme proposed by Lechner, Hauke, and Zoller (LHZ) \cite{lechner2015quantum}, in which a pair of interacting logical qubits is replaced by a single physical qubit with an appropriate constraint among the latter qubits being introduced to reproduce the original Hamiltonian. The classical part of the Hamiltonian in the LHZ scheme is composed of local longitudinal fields representing the original problem Hamiltonian and nearest-neighbor four-body interactions for the constraint mentioned above. This scheme has been studied toward practical implementation \cite{leib2016transmon,nigg2017robust,chancellor2017circuit,puri2017engineering,puri2017quantum,zhao2018two,goto2019quantum,kanao2020high}.
It is nevertheless noted that the standard minor embedding is in some cases more advantageous than the LHZ scheme according to numerical simulations in Ref.~ \cite{albash2016simulated}.
It was also reported that the performance of QA under the LHZ scheme can be enhanced by the counter diabatic control of system parameters \cite{hartmann2019rapid} and inhomogeneous driving of the transverse field \cite{hartmann2019quantum}. In addition, a mean-field analysis assuming adiabatic parameter control shows that nonlinear driving of the coefficient of the constraint term is effective to avoid a problematic first-order phase transition \cite{susa2020performance}.

Theoretical analyses of QA have traditionally been carried out in the framework of adiabatic computation, in which the system stays in the instantaneous static ground state during the annealing process. However, it has gradually been recognized that investigating of diabatic (i.e., nonstatic) processes may at least be equally important. The reasons are, first, a real device usually operates away from the adiabatic regime, second, a shorter annealing time would be beneficial to avoid decoherence,  third, repeating short annealing processes may sometimes lead to better results than a single long adiabatic process, and last, diabatic annealing cannot be efficiently simulated classically generally, which may lead to a quantum speedup.  See Ref.~\cite{Crosson2020} for a review. 
\footnote{Reference~\cite{callison2020energetic} developed a related idea, but their arguments do not apply directly to the present case.}

One of the promising approaches to diabatic annealing is a variational method recently proposed by Matsuura {\it et al.} \cite{matsuura2020vanqver,matsuura2020variationally}, in which one iteratively improves the time dependence of coefficients in the Hamiltonian to minimize the energy expectation value. This is an idea inspired by a variational quantum eigensolver \cite{peruzzo2014variational,mcclean2016theory}, a quantum-classical hybrid algorithm, to find the eigenstate of a quantum system in the noisy near-term quantum device.
They applied the variational method to a small-scale quantum chemistry problem and an Ising model with frustration and concluded that much shorter annealing times (strongly diabatic processes) are sufficient to achieve satisfactory results.

Given these developments, we follow Refs.~\cite{matsuura2020vanqver,matsuura2020variationally} and adapt the method to the LHZ scheme toward performance enhancement under diabatic conditions. We numerically show that the variational optimization of the annealing schedule of the constraint term in the LHZ scheme leads to a significant improvement of performance measured by the residual energy and ground-state fidelity using small-scale examples of a simple ferromagnet and the spin-glass problem.

This paper is organized as follows. Section \ref{sec:review} introduces the LHZ scheme, and Sec. \ref{sec:vsqs} explains the variational algorithm applied to the LHZ scheme. Numerical results are presented in Sec. \ref{sec:result} and discussions are given in Sec. \ref{sec:discussion}. 
\vspace{-1.5ex}
\section{Lechner-Hauke-Zoller scheme}
\label{sec:review}

We first recapitulate the LHZ scheme \cite{lechner2015quantum} to set the stage for developments in the following sections. The Hamiltonian describing QA has traditionally been chosen as
\begin{align}
\hat{H}(t) = \qty(1-\frac{t}{T})\sum_{i=1}^N \hat{\sigma}_i^x + \frac{t}{T}\hat{H}_{P},
\label{eq:TFIM_original}
\end{align}
where $\hat{\sigma}_i^{x}$ represents the $x$ component of the Pauli operator of qubit (spin) $i$. The total annealing time is $T$. 
The first term in the Hamiltonian is the transverse field, and its coefficient is decreased from 1 ($t=0$) to 0 ($t=T$). The second term is the problem Hamiltonian $\hat{H}_{P}$ for the cost function of combinatorial optimization problem written in terms of the $z$ component of the Pauli operator. More explicitly, the problem Hamiltonian is often the fully connected Ising model,
\begin{align}
\hat{H}_{\text{Ising}} = -\sum_{i<j}J_{ij}\hat{\sigma}_i^z\hat{\sigma}_j^z-\sum_{i=1}^N h_i \hat{\sigma}_i^z.
\label{eq:Ising_original}
\end{align}

The LHZ scheme converts the all-to-all connectivity in the above Hamiltonian to local terms. The interaction $J_{ij}\hat{\sigma}_i^z\hat{\sigma}_j^z$, where $\{\hat{\sigma}_i^z\}$ are logical qubits, is replaced by a longitudinal-field term $J_k \sigma_k^z$, where $\sigma_k^z$ without a hat is for a physical qubit, and four-body interactions among neighboring physical qubits are added as a constraint to recover the original Hamiltonian. The resulting Hamiltonian is written as
\begin{align}
\label{eq:lhz}
H_{\text{LHZ}} = -\sum_{k=1}^K J_k \sigma_k^z - C\sum_{l=1}^L \sigma_{(l,n)}^z\sigma_{(l,w)}^z\sigma_{(l,s)}^z\sigma_{(l,e)}^z,
\end{align}
where $K=N(N-1)/2$ is the total number of physical qubits and $L = (N-1)(N-2)/2$ is the number of constraints. As long as the product of four qubits takes the value $+1$, $H_{\text{LHZ}}$ is equivalent to $\hat{H}_{\text{Ising}}$ of Eq.~(\ref{eq:Ising_original}). A transverse field is added to the above Hamiltonian in the usual way as in Eq. (\ref{eq:TFIM_original}), with $H_{\text{LHZ}}$ replacing $\hat{H}_{P}$. Reference \cite{lanthaler2020minimal} reports that an appropriate value of $C$ can be determined according to the class of the problem and the system size. Reference ~\cite{susa2020performance} develops a static analysis and finds it better to control the value of $C$ differently from those of other coefficients to avoid a first-order phase transition. In the present paper, we introduce time dependence in the coefficient as $C(t/T)$ and iteratively update its time functional form to maximize the performance. 

\begin{figure*}[htb]
\centering
\subfigure{
\includegraphics[height = 4.4cm]{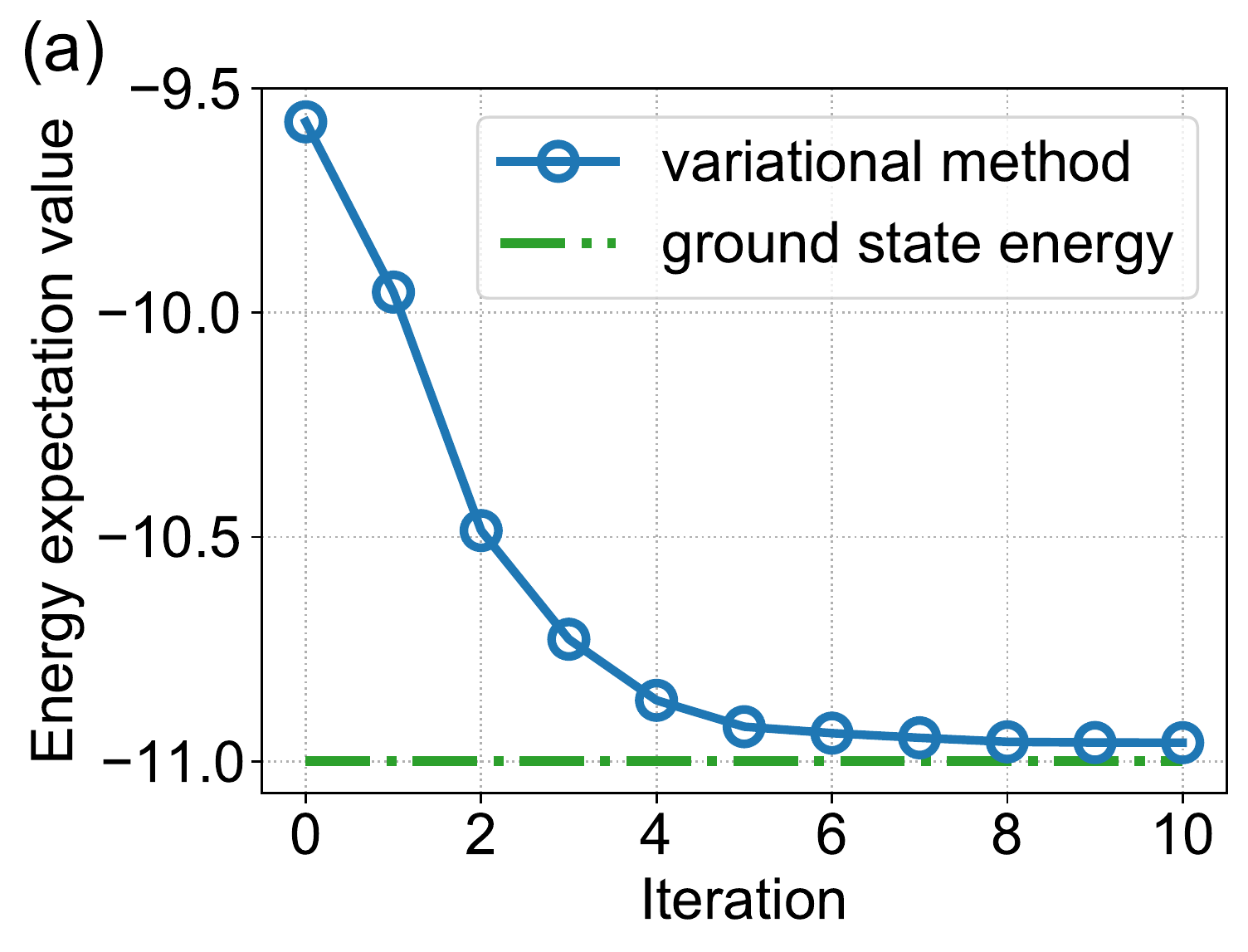}
\label{fig2a}
}
\subfigure{
\includegraphics[height = 4.4cm]{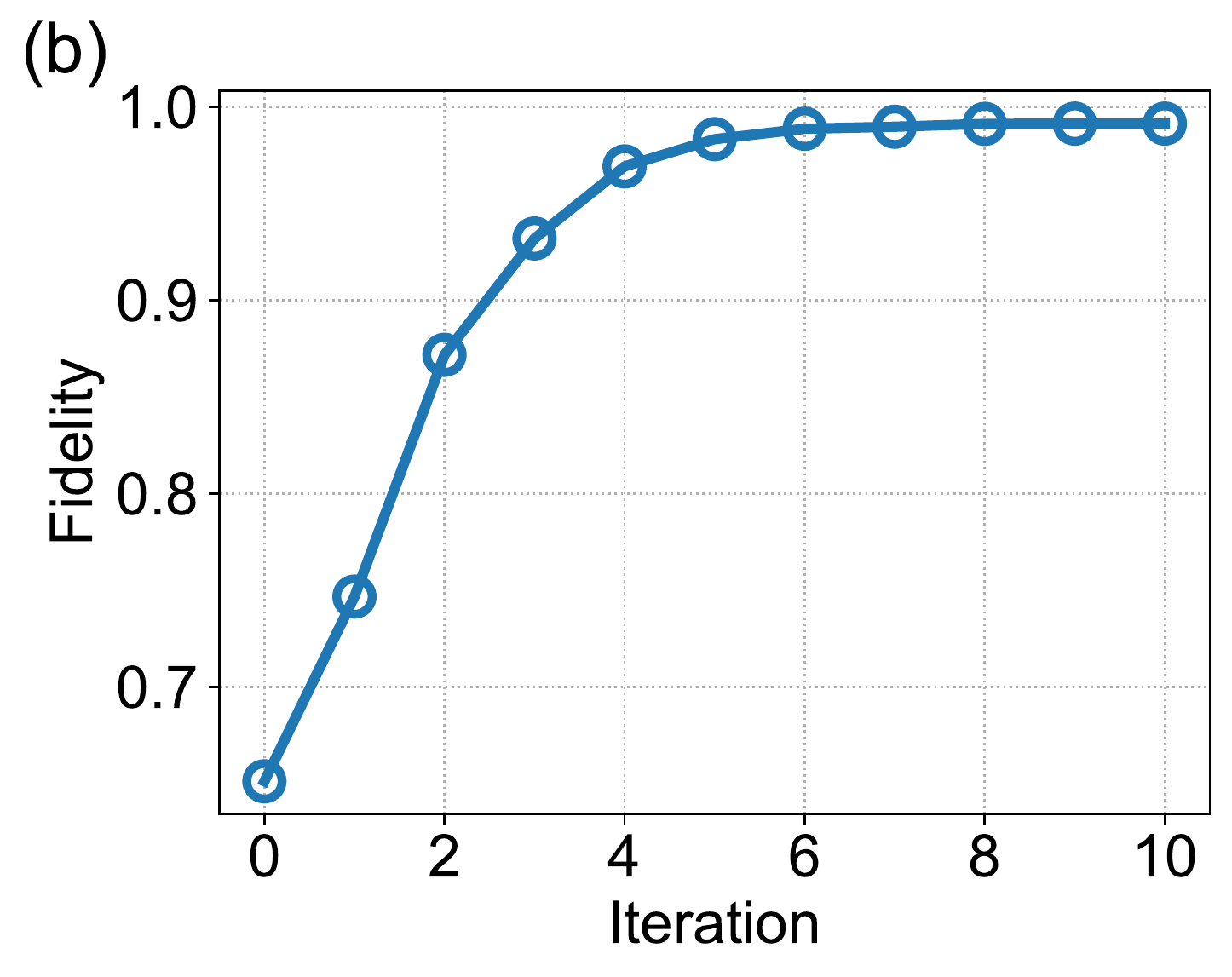}
\label{fig2b}
}
\subfigure{
\includegraphics[height = 4.4cm]{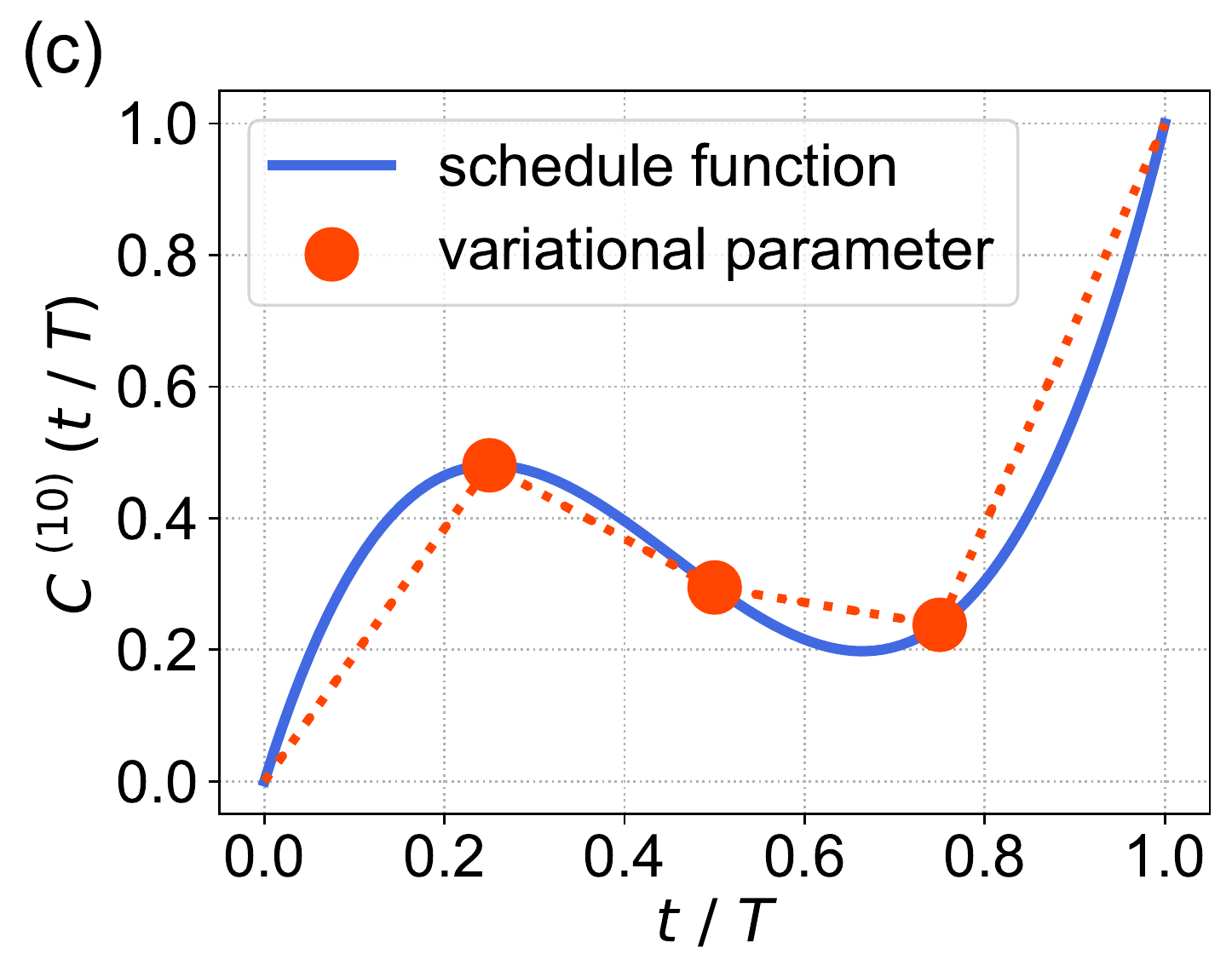}
\label{fig2c}
}
\caption{Numerical results for a ferromagnetic model with  $J_k=0.5$. (a) and (b) show the energy expectation value $\langle E \rangle^{(m)}$ and the final ground-state fidelity $F^{(m)}$ as functions of the iteration number $m$, respectively. (c) The variational parameters (red circles) and its polynomial fit (blue solid curve) after ten iterations.}
\label{fig2}
\end{figure*}
\vspace{-1.5ex}
\section{Variational algorithm}
\label{sec:vsqs}

We adapt the variational method of Ref.~\cite{matsuura2020variationally} to the determination of the optimal time dependence of the coefficient $C(t/T)$. The total Hamiltonian is now
\begin{align}
H(t/T) = & \qty(1-\frac{t}{T})\sum_{k=1}^K \sigma_k^x - \frac{t}{T}\sum_{k=1}^K J_k \sigma_k^z \notag \\
&- C(t/T) \sum_{l=1}^L \sigma_{(l,n)}^z \sigma_{(l,w)}^z \sigma_{(l,s)}^z\sigma_{(l,e)}^z.
\end{align}
For practical purposes of variational optimization, we discretize the time as $t/T=n/(S+1)$ ($n=0, 1,2,\cdots, S+1$) and represent the function $C(t/T)$ at discrete points $c_n=C\mathbf{(}n/(S+1)\mathbf{)}$ with the boundary condition $c_0=C(0)=0$ and $c_{S+1}=C(1)=1$.  The principle of variational optimization is to iteratively change the value of $c_n$ toward minimization of the expectation value of the final Hamiltonian $H(1)$ by repeating the process of short-time quantum annealing. The value of the coefficient at the $m\text{th}$ iteration will be denoted as $c_n^{(m)}$. The update of the value of $c_n$ is performed on a classical computer.
After each step of classical update, we fit a polynomial of degree $D$,
\begin{align}
C^{(m+1)}(t/T) = \sum_{d=1}^D a_d (t/T)^d,\quad a_1 = 1-\sum_{d=2}^D a_d,
\end{align}
to the discrete values of $\{c_n^{(m+1)}\}$ and run the next annealing process using this $C^{(m+1)}(t/T)$ following the continuous-time Schr\"odinger dynamics, a quantum process. This is thus a classical-quantum hybrid algorithm. Figure~\ref{fig1} summarizes the whole algorithm.
\vspace{-1.5ex}
\section{Numerical result}
\label{sec:result}
We now show the results obtained by numerically solving the Schr\"odinger equation.  The Broyden-Fletcher-Goldfarb-Shanno algorithm was used as the classical optimizer \cite{broyden1970convergence,fletcher1970new,goldfarb1970family,shanno1970conditioning,fletcher1987practical}.
The initial parameters are chosen as $c_{n}^{(0)}=n/(S+1)$ and $C^{(0)}(t/T)=t/T$. The number of physical spins is $K=10$ and we iterate the loop ten times.

\subsection{Ferromagnetic model}

First, we test the framework by the simplest ferromagnetic model with $J_k = 0.5$ with the results in Fig. \ref{fig2}.  The number of the variational parameters is $S = 3$, and the order of fitting polynomial is $D=4$. The annealing time is fixed to $T=10$. Figures \ref{fig2a} and \ref{fig2b} are for the energy expectation value $\langle E \rangle^{(m)}$ and the fidelity $F^{(m)}=|\langle\phi^{(m)}|\psi_{\textrm{GS}}\rangle|^2$, respectively,  with $|\phi^{(m)}\rangle$ the obtained ground state after $m$ iterations and $\ket{\psi_{\textrm{GS}}}$ the true ground state.
The data point at $m=0$ corresponds to the linear schedule function $C^{(0)}(t/T)$, where it is clear that the system remains far from the ground state, i.e., far from adiabaticity under the linear annealing schedule with $T=10$.
By iterating the loop, the energy expectation value approaches the true ground-state energy  and correspondingly the fidelity is improved toward 1. Figure \ref{fig2c} displays the time dependence of the final variational parameters $\{c_{n}^{(10)}\}$ and the corresponding schedule $C^{(10)}(t/T)$.

\begin{figure}[b]
\centering
\includegraphics[height =4cm]{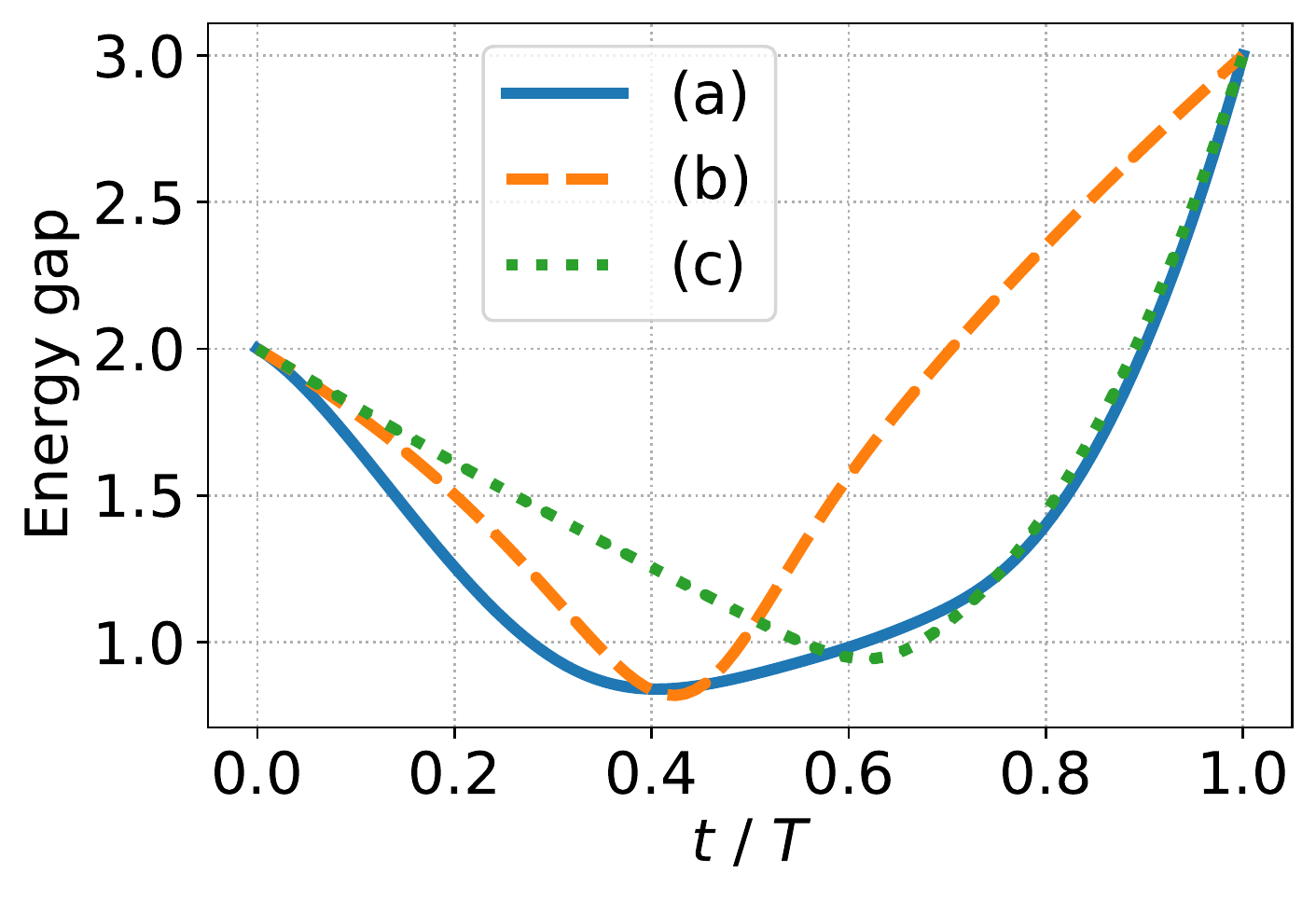}
\caption{Energy gap between the ground state and the first-excited state as a function of  time  for three schedules, (a) $C(t/T)=C^{(10)}(t/T)$, (b) $C(t/T) = t/T$, and (c) $C(t/T) = (t/T)^5$.}
\label{fig3}
\end{figure}

In an effort to clarify the significance of the obtained annealing schedule, we have calculated the energy gap between the instantaneous ground state and the first-excited state as a function of time. The result is displayed in Fig. \ref{fig3}. We also plot the corresponding energy gaps for the linear function $C(t/T) = t/T$ and a monotonically increasing nonlinear function $C(t/T) = (t/T)^5$, the latter of which avoids the first-order transition in the static (adiabatic) phase diagram \cite{susa2020performance}. The minimum gap of the variationally optimal schedule [Fig. \hyperref[fig3]{3(a)}] is almost the same as the simple linear case [Fig. \hyperref[fig3]{3(b)}], and the nonlinear schedule [Fig. \hyperref[fig3]{3(c)}] has a little larger gap, possibly reflecting the avoidance of the first-order transition. This latter slightness of the increase of the gap would come from the smallness of the system size.  These results may suggest the difficulty to predict dynamical, diabatic properties from the energy gap. We should nevertheless be cautious since the the data on the energy gap may reflect finite-size effects.

\begin{figure*}[t]
\centering
\subfigure{
\includegraphics[height = 4.5cm]{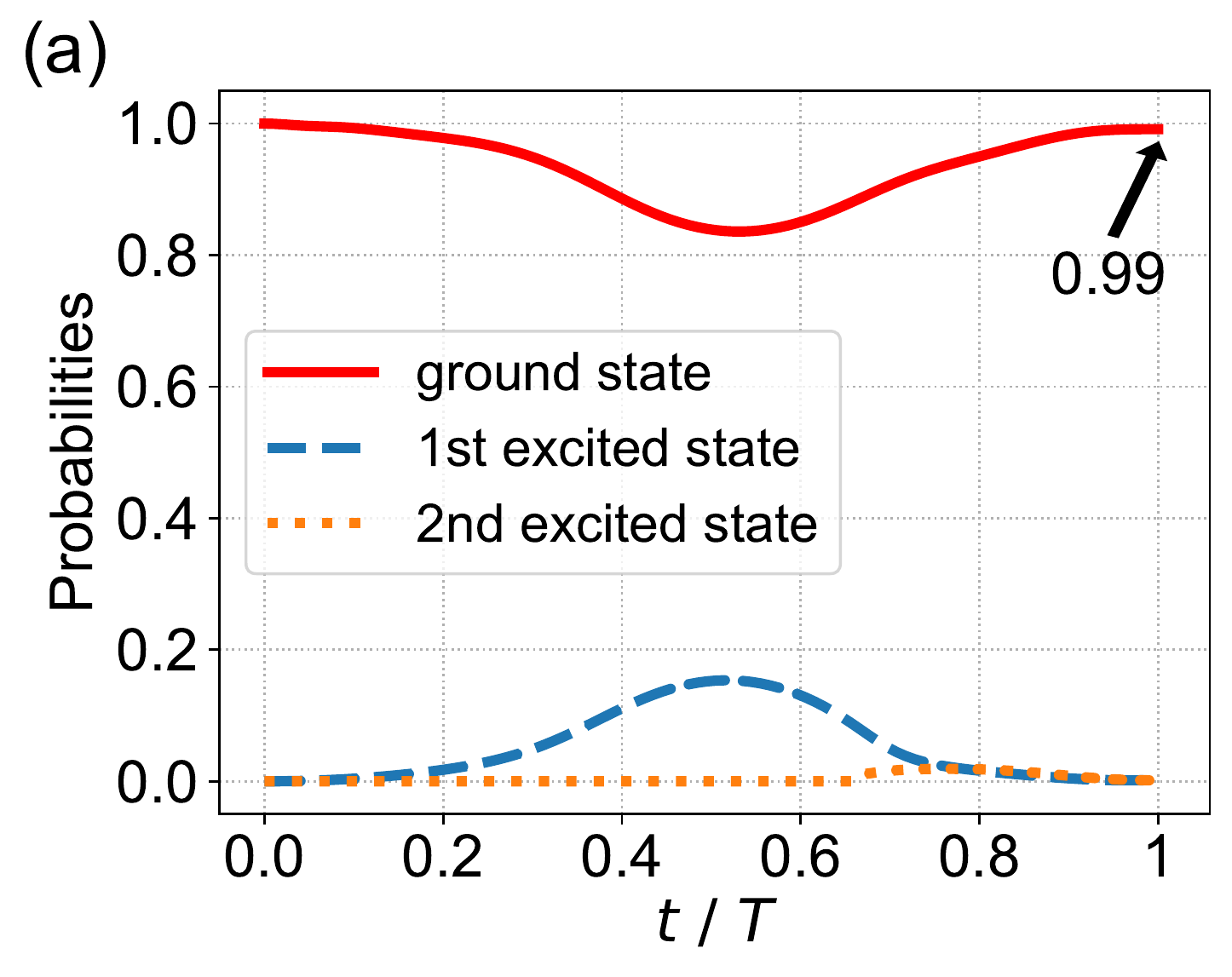}
\label{fig4a}
}
\subfigure{
\includegraphics[height = 4.5cm]{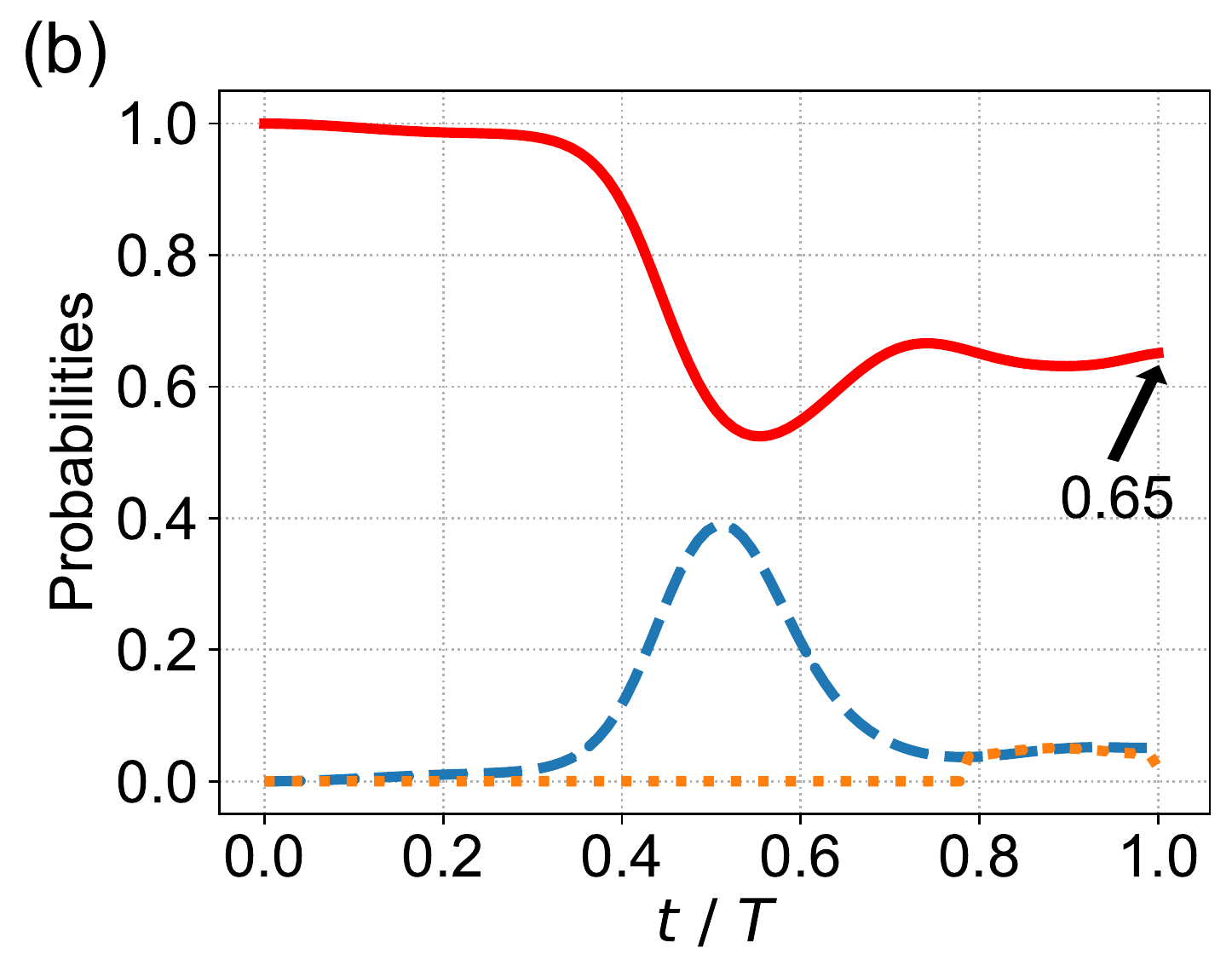}
\label{fig4b}
}
\subfigure{
\includegraphics[height = 4.5cm]{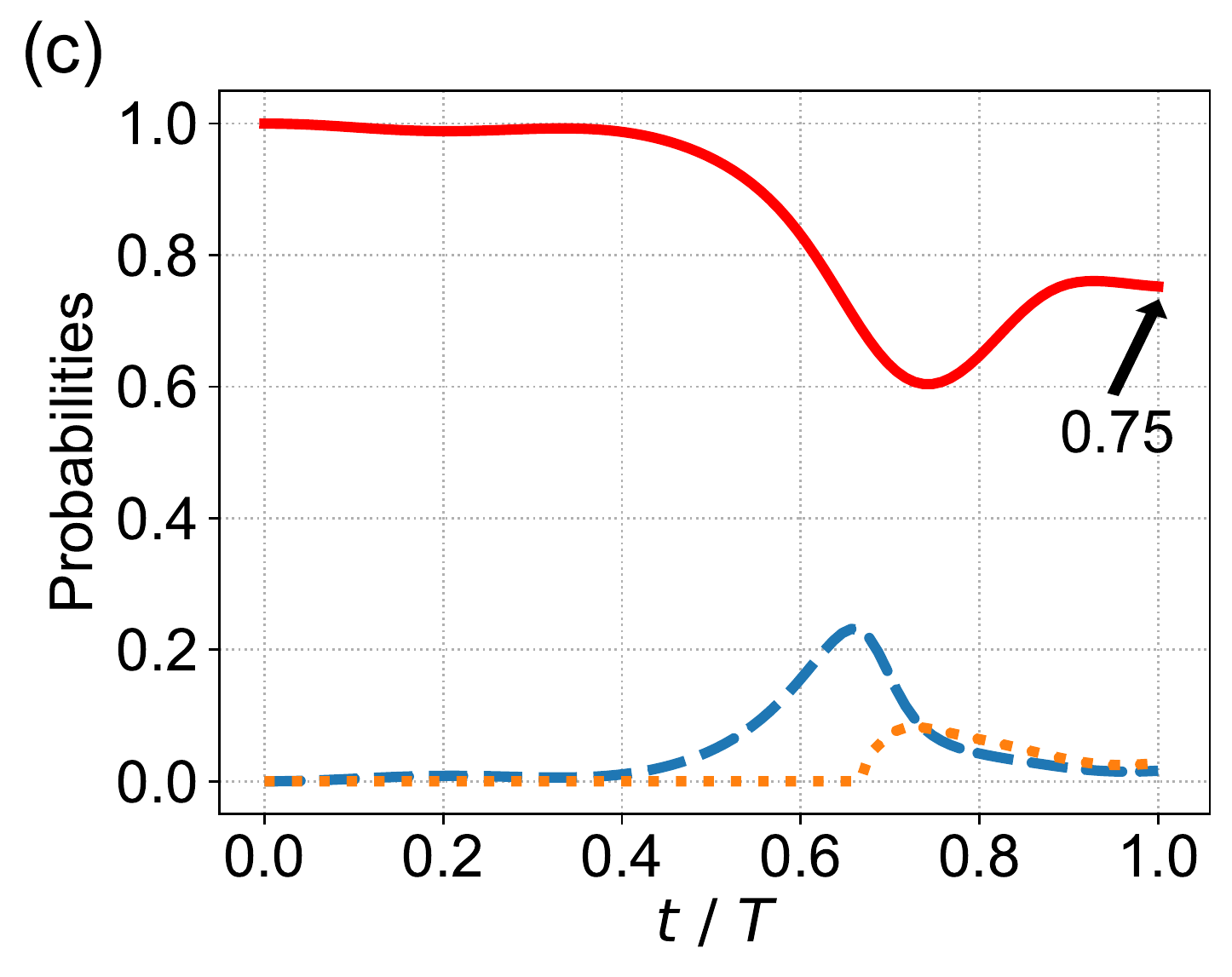}
\label{fig4c}
}
\caption{Probabilities of three lowest eigenstates as functions of time for the three schedules in Fig. \ref{fig3}. (a)--(c) correspond to the annealing schedules (a)--(c) in Fig.~\ref{fig3}, respectively. Curves in (b) and (c) are distinguished by the same color code and line type as in (a).}
\label{fig4}
\end{figure*}

\begin{figure*}[htb]
\centering
\subfigure{
\includegraphics[height =4.3cm]{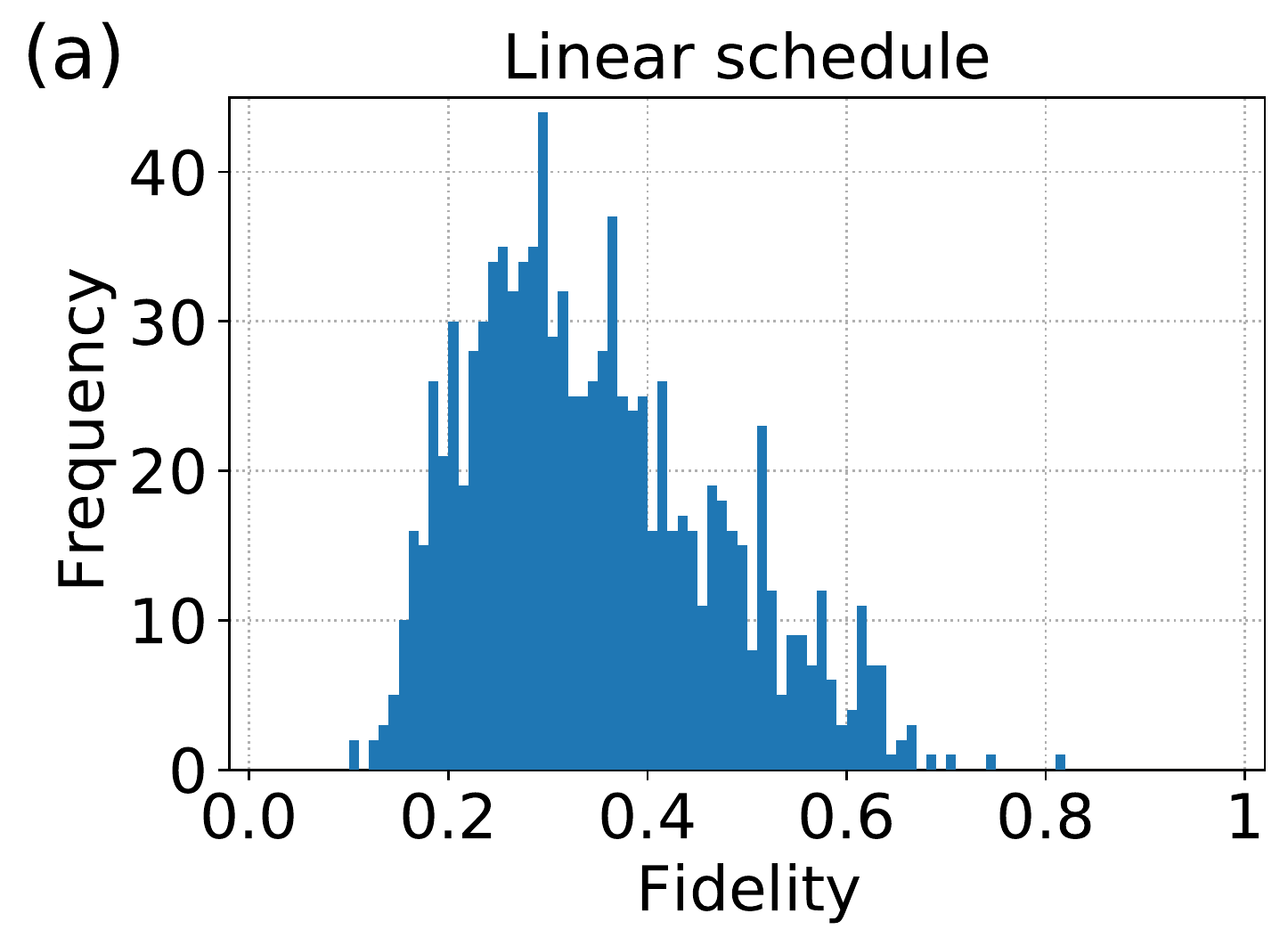}
\label{fig5a}
}
\subfigure{
\includegraphics[height =4.3cm]{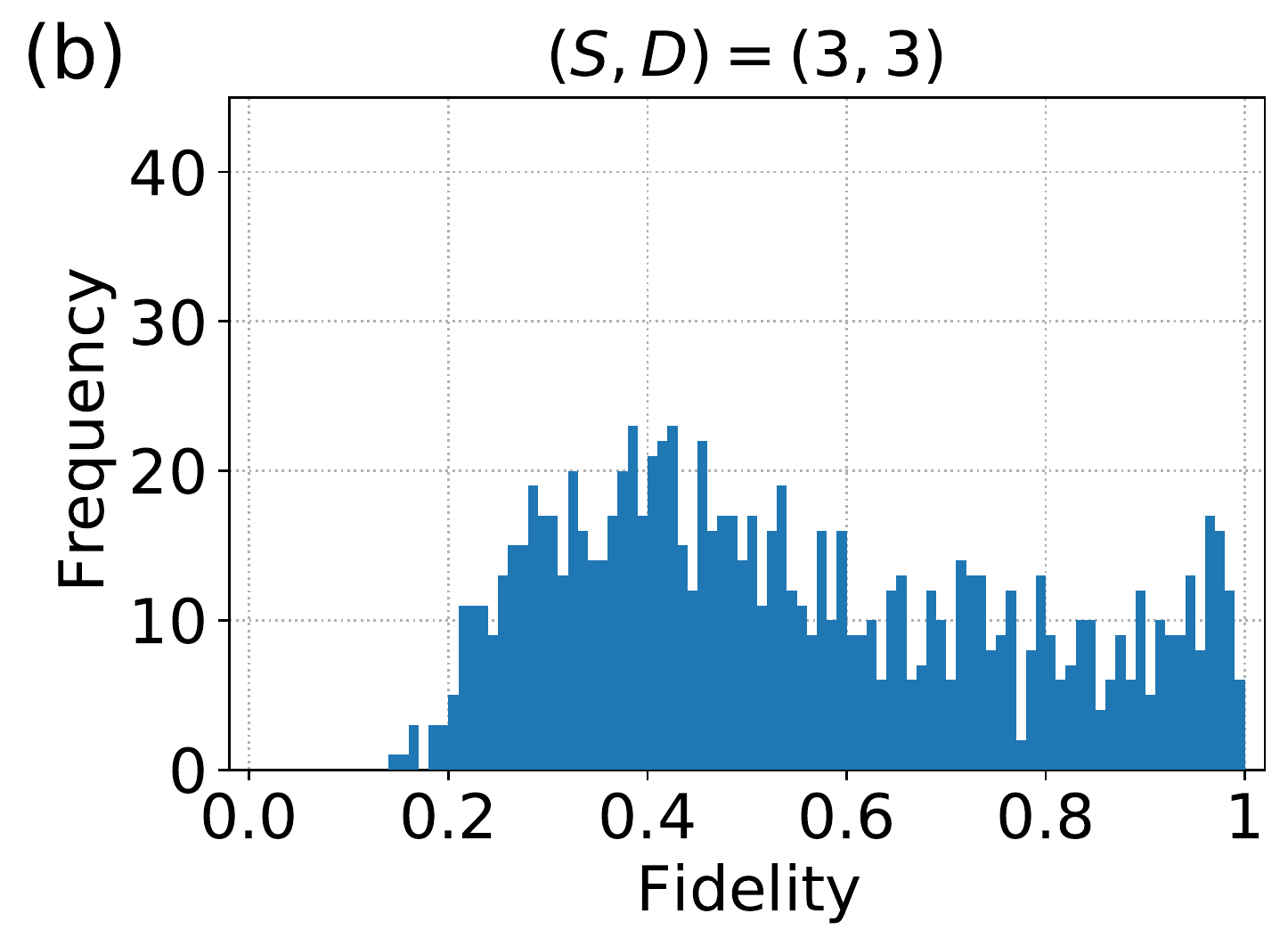}
\label{fig5b}
}\\
\subfigure{
\includegraphics[height =4.3cm]{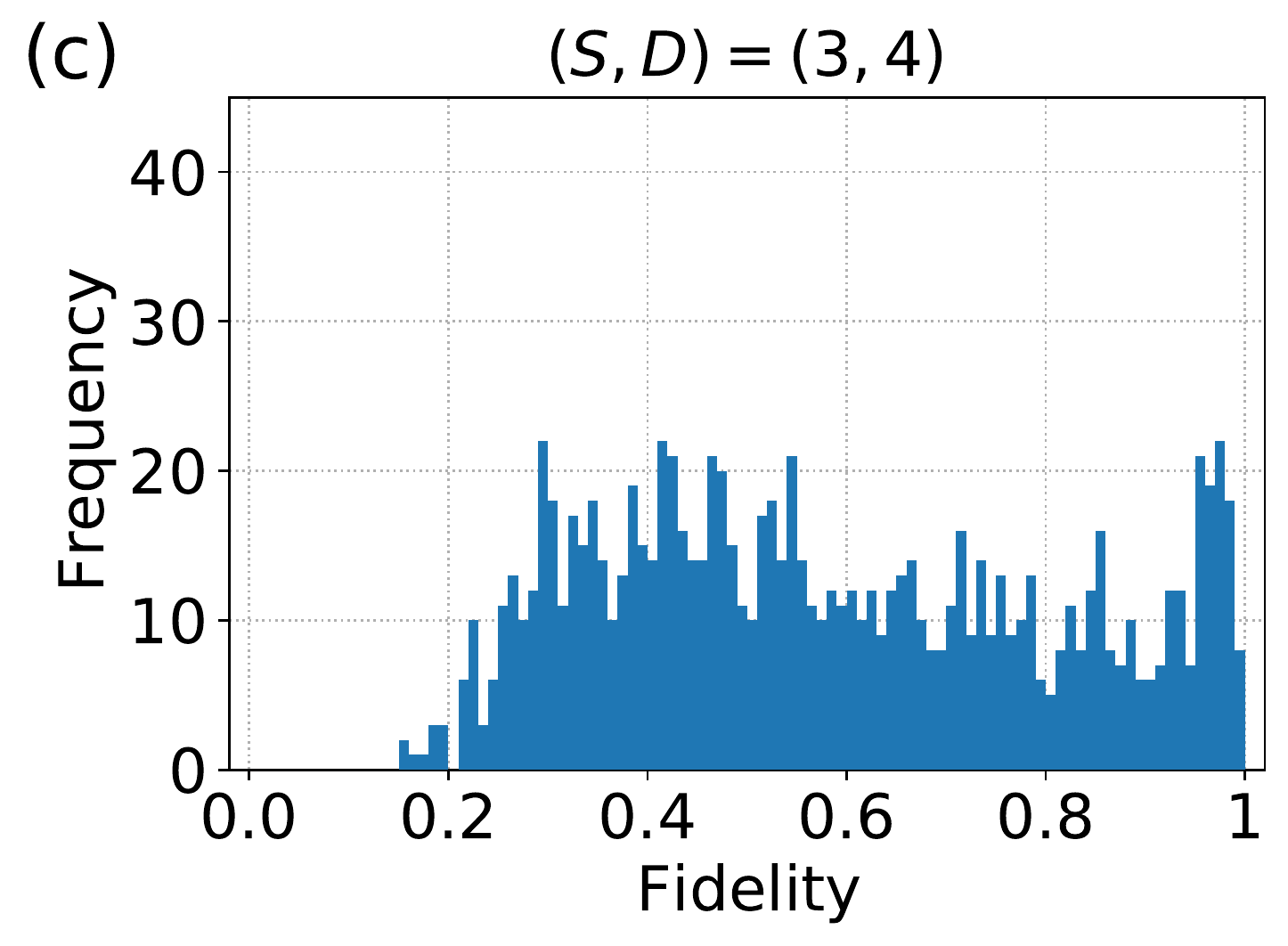}
\label{fig5c}
}
\subfigure{
\includegraphics[height =4.3cm]{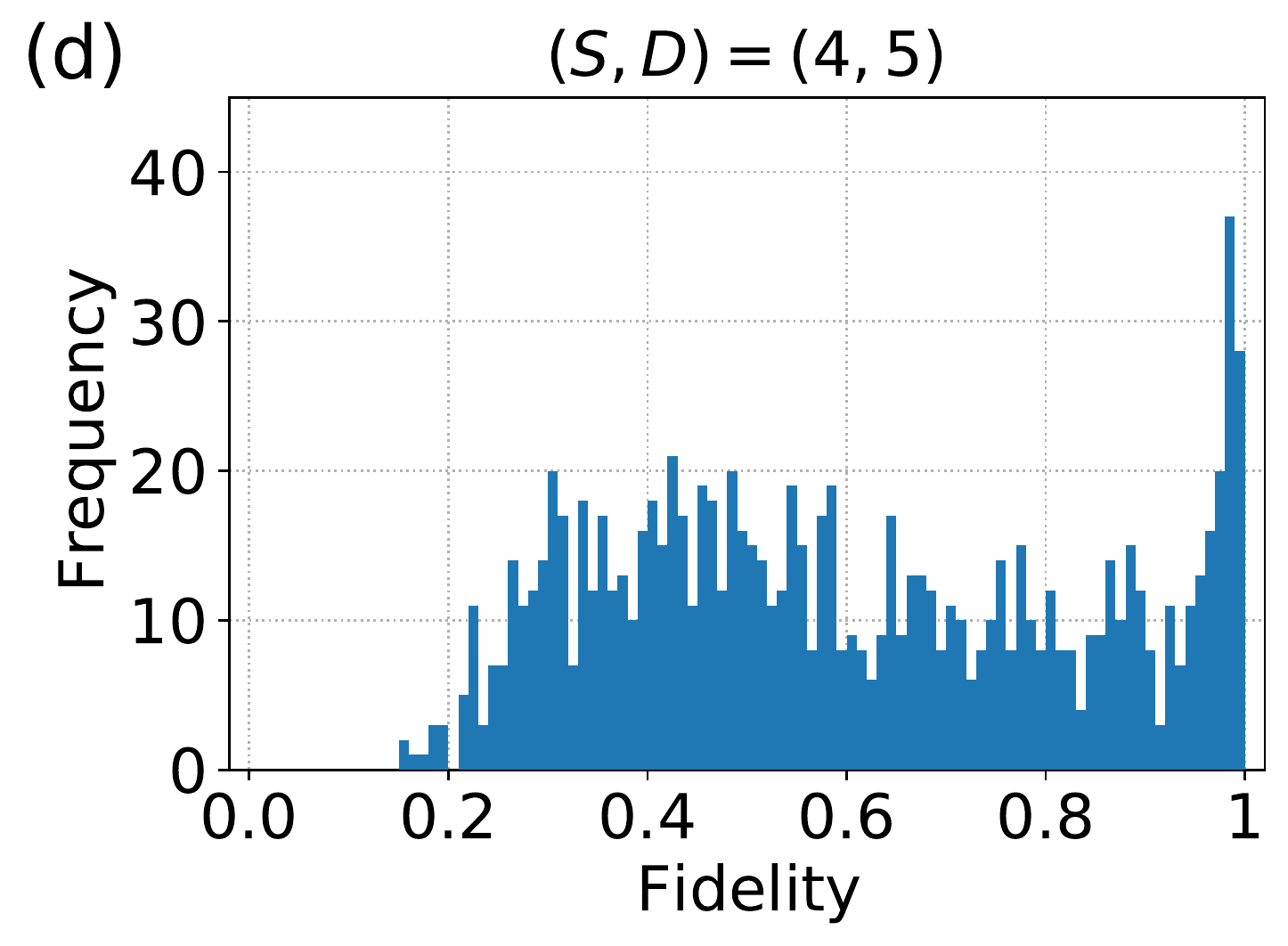}
\label{fig5d}
}
\caption{Histogram of the ground-state fidelity of 1000 random instances of the spin-glass problem. We employ the simple linear function $C(t/T) = t/T$ in (a) and the variational ones in (b)--(d), where the values of $S$ and $D$ are chosen as displayed. In all panels, we use the same instances of randomness.}
\label{fig5}
\end{figure*}

We have further calculated the probabilities of the three lowest instantaneous eigenstates as functions of time for the three schedules used in Fig. \ref{fig3}.

As seen in Fig.~\ref{fig4a}, the ground-state probability drops around the middle of annealing, and but finally increases toward 1 for the variationally obtained schedule.  On the other hand, as observed in Figs.~\ref{fig4b} and \ref{fig4c}, the probability stays far from 1 after a drop for $C(t/T) = t/T$ and $C(t/T) = (t/T)^5$. It is obvious that the variational algorithm succeeds to keep the system close to the instantaneous ground state at least for the present small-size problem with a short annealing time, where the traditional method fails.

\begin{figure*}[tb]
\centering
\subfigure{
\includegraphics[height =4.8cm]{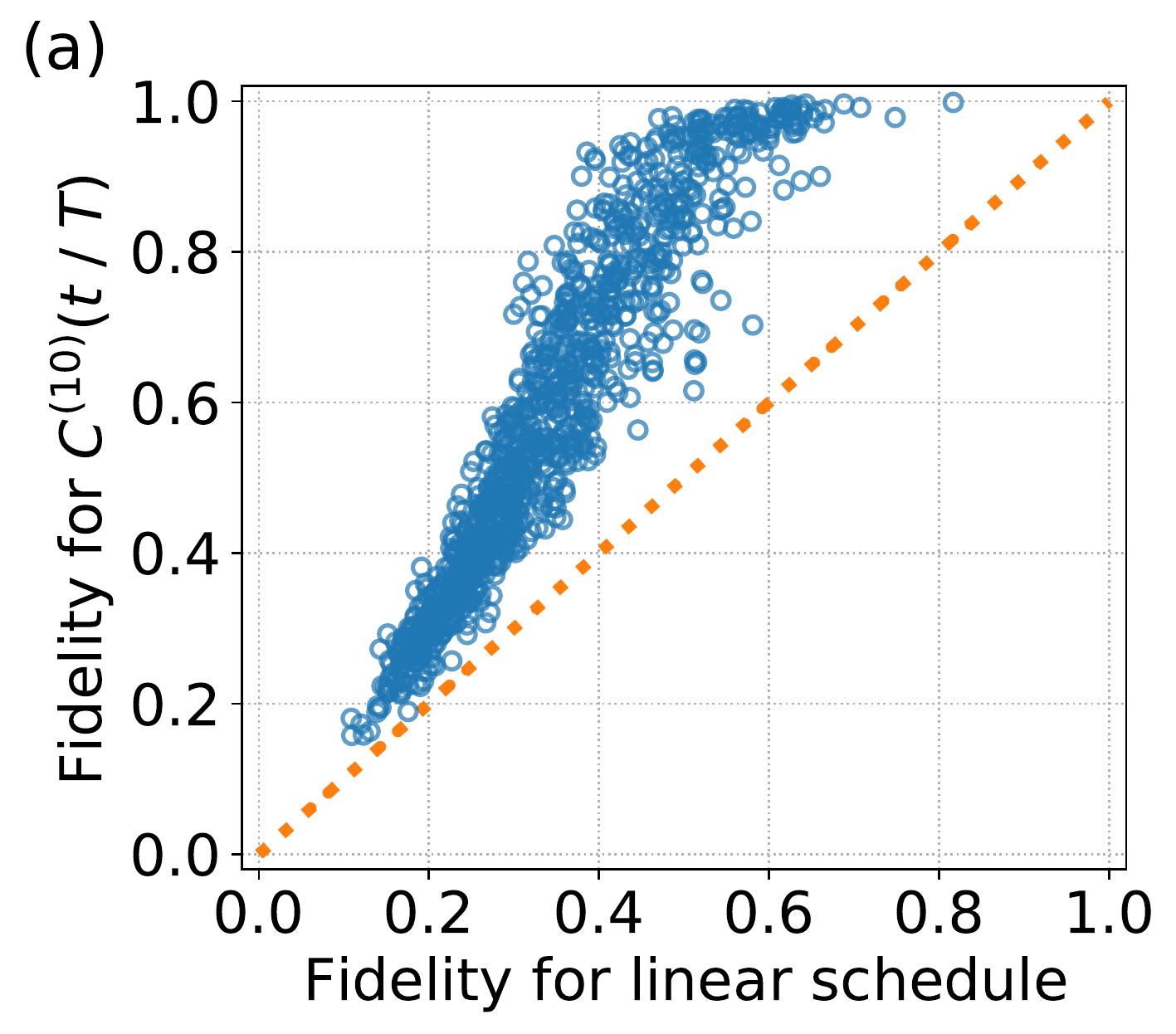}
\label{fig6a}
}
\subfigure{
\includegraphics[height =4.8cm]{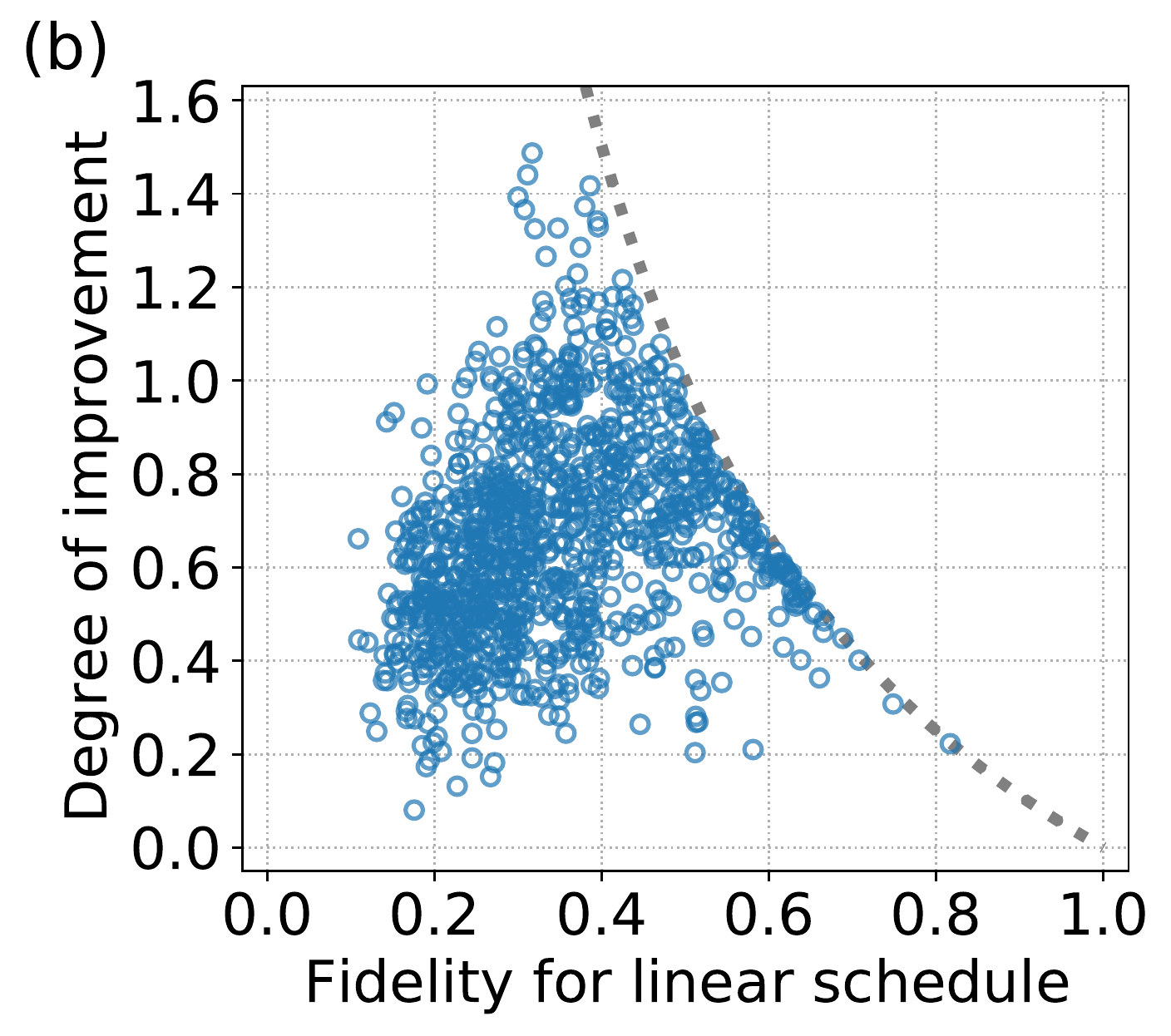}
\label{fig6b}
}
\subfigure{
\includegraphics[height =4.3cm]{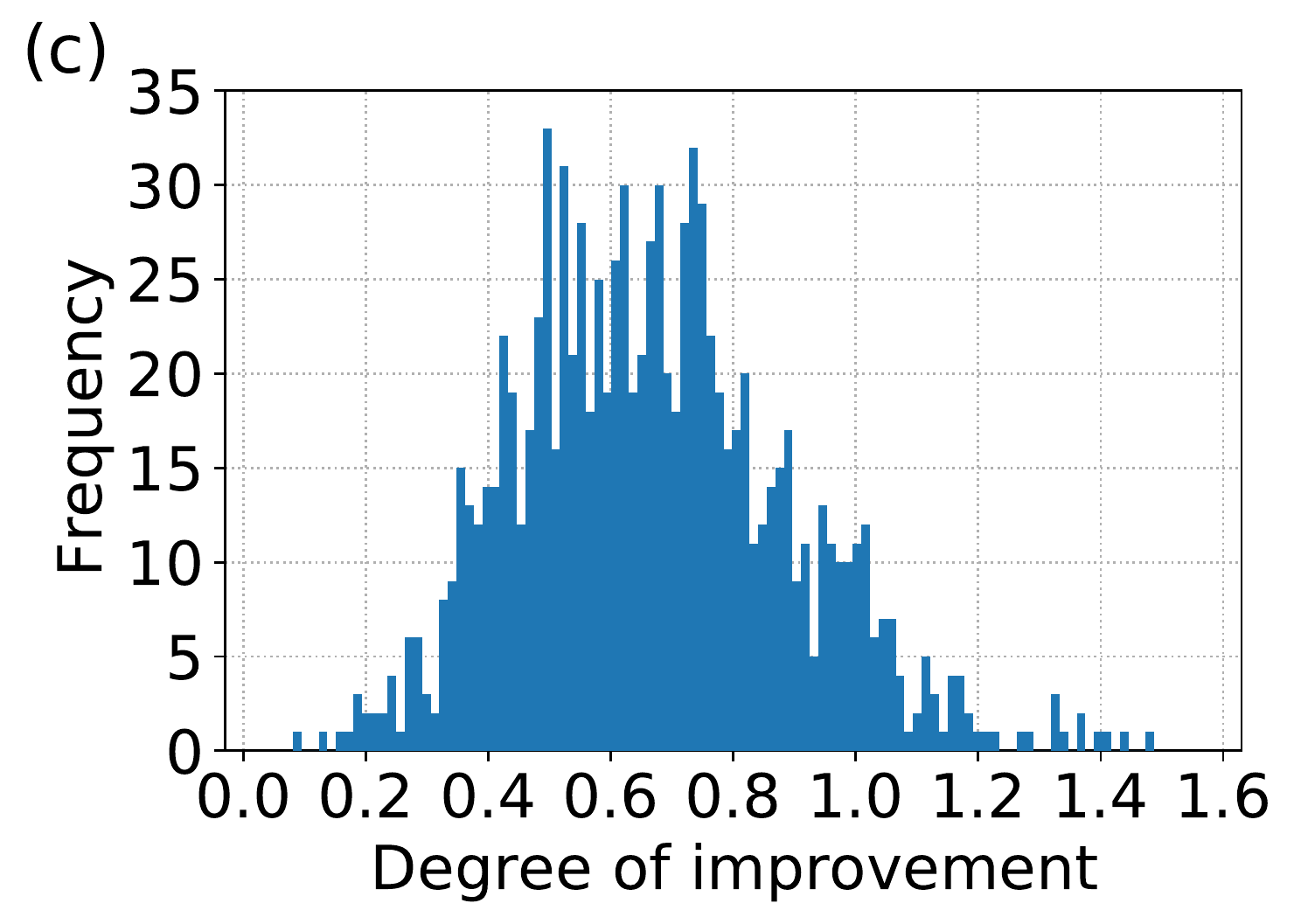}
\label{fig6c}
}
\subfigure{
\includegraphics[height =4.8cm]{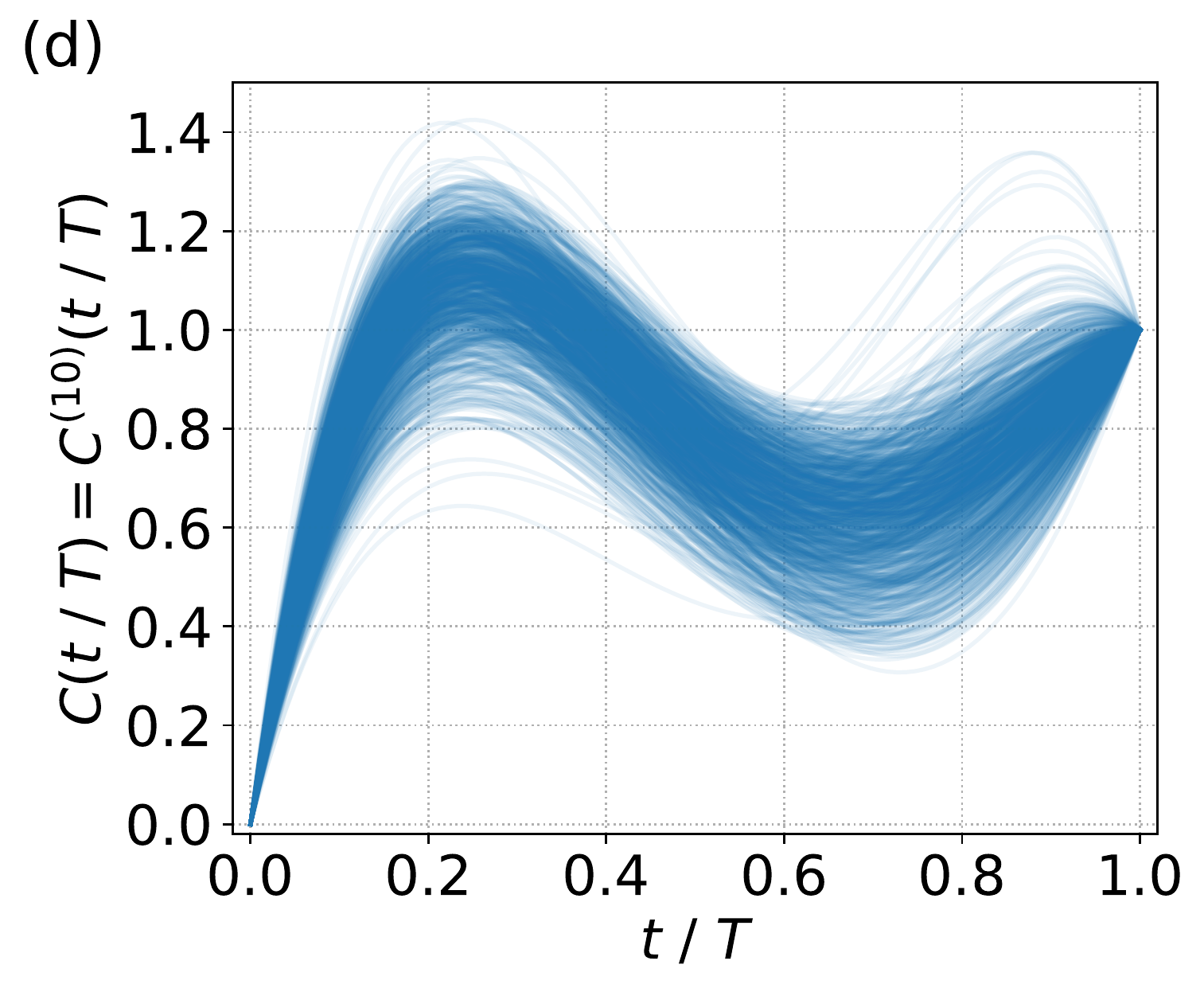}
\label{fig6d}
}
\subfigure{
\includegraphics[height =4.6cm]{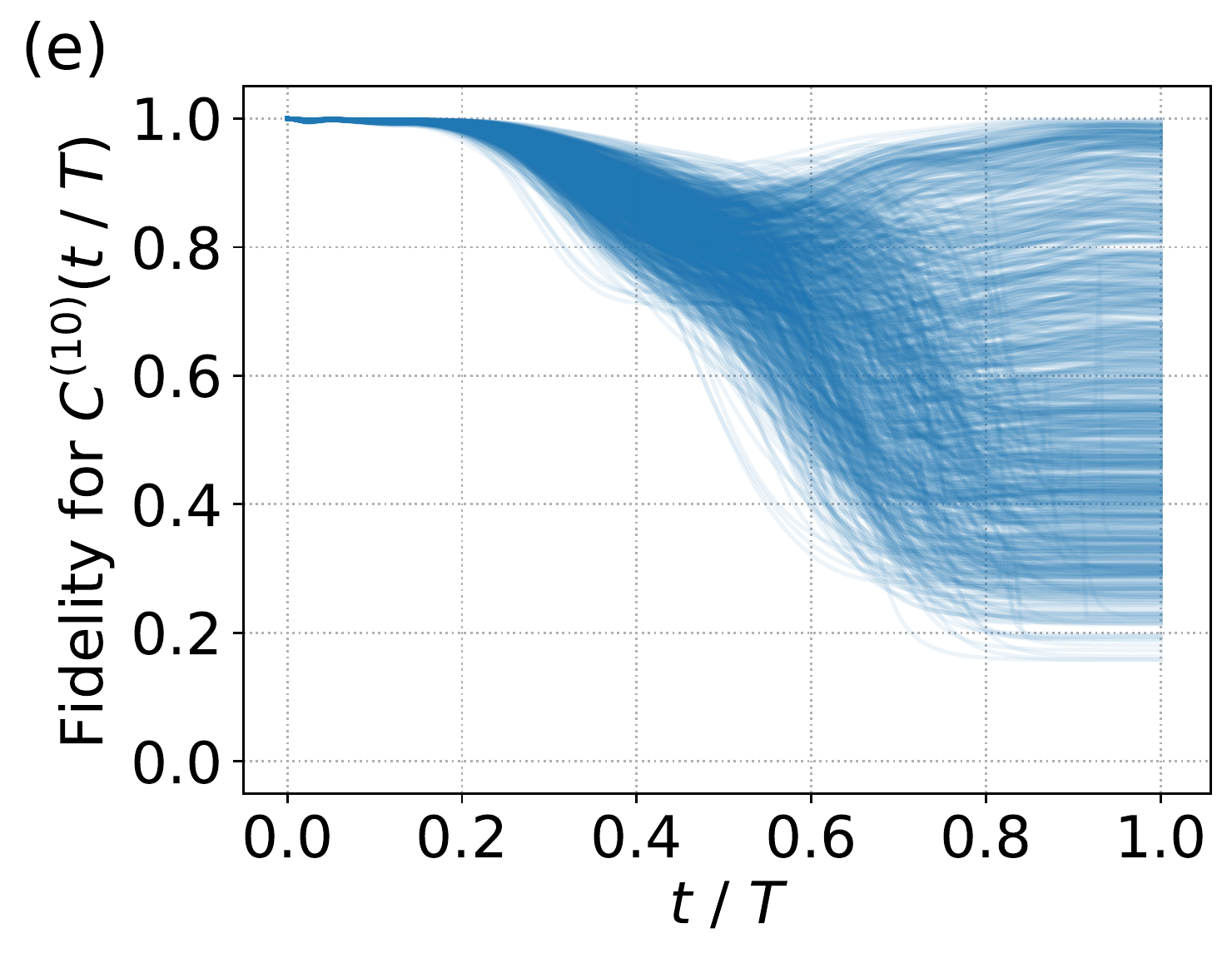}
\label{fig6e}
}
\caption{(a) Scatter plot of the fidelity in the variational method against the one for the linear schedule for 1000 random instances of the spin-glass problem. (b) Scatter plot of the degree of improvement of the fidelity by the variational method against the fidelity in the linear schedule. The gray dashed curve indicates the upper bound of the fidelity improvement. (c) Histogram of the normalized degree of improvement. (d) Optimized schedules for 1000 random instances. (e) The fidelity of the instantaneous ground state as a function of time for 1000 random instances with the optimal schedules for each instance shown in (d). In each plot, we use the same instances considered in Fig. \ref{fig5} and fix the parameters $(S,D) = (3,4)$.}
\label{fig6}
\end{figure*}

\vspace{-3ex}
\subsection{Spin-glass problem}

Let us next consider the spin-glass problem with $J_k$ uniformly chosen from the range $-0.5<J_k<0.5$. This independent random choice of $J_k$ for a local field under the constraint of a four-body interaction realizes the spin-glass problem in the LHZ scheme \cite{lechner2015quantum}. It is useful to focus on the final ground-state fidelity because there exist many excited states with the energy very close to the ground-state energy in the spin-glass problem, which makes it difficult to measure the performance by the residual energy. 
\vspace{-0.5ex}

Figure \ref{fig5} shows the histogram of the ground-state fidelity obtained by running the variational algorithm for 1000 random instances of $\{J_k\}$ with a fixed annealing time $T = 20$.
Four annealing schedules have been studied, the linear function [Fig. \hyperref[fig5a]{5(a)}], and three variationally optimized functions with parameters $(S,D) = (3,3)$ [Fig. \hyperref[fig5b]{5(b)}], $(S,D) = (3,4)$ [Fig. \hyperref[fig5c]{5(c)}], and $(S,D) = (4,5)$ [Fig. \hyperref[fig5d]{5(d)}]. We see in Fig. \ref{fig5a} that with the linear schedule the fidelity falls between 0.1 and 0.6 for most instances. The system did not achieve the ground state (fidelity close to one) for any example in this case.
The other histograms for the variational method show significant improvements. It is observed that larger numbers of variational parameters $S$ and $D$ naturally lead to better results.
\vspace{-0.5ex}

Figure \ref{fig6a} is a scatter plot of fidelity for the optimized schedule with $(S,D) = (3,4)$ versus the linear schedule. 
Figure \ref{fig6b} shows the degree of improvement, defined as the relative difference between fidelity for the optimized schedule and for the linear one which is $(\text{vertical axis} - \text{horizontal axis})/(\text{horizontal axis})$ in Fig.~\ref{fig6a}, for each data point in Fig.~\ref{fig6a}. The dotted curve displays $(1- \text{horizontal axis})/(\text{horizontal axis})$ representing the upper bound of the fidelity improvement and the data points near the dotted curve show the cases where the variational algorithm improves the fidelity to 1. It is observed that an easy instance for the linear schedule is more likely to benefit from the variational  method than a difficult instance. The ground-state fidelity is improved mostly over 40\% regardless of the difficulty. 
Figure~\ref{fig6c} is the histogram of the degree of improvement shown in Fig.~\ref{fig6b}.  Most of the instances have values between 0.4 and 0.8 with scattered outliers. The lower tail of the histogram in Fig.~\ref{fig6c}  may correspond to the final states with the energy close to the ground state but the fidelity or the Hamming distance far away. Figure~\ref{fig6d} is for optimized schedules, most of which look similar to the ferromagnetic case in Fig. \ref{fig2c}.
Similarly to Fig.~\ref{fig4}, we investigated the fidelity of the instantaneous ground state for 1000 random instances under the optimized schedule shown in Fig. \ref{fig6d} and the result is plotted in Fig. \ref{fig6e}. We find that for all instances the fidelity stays close to 1 until $t/T = 0.25$, where the optimized schedules in Fig.~\ref{fig6d} reach their peak.
This result may suggest that the system is prepared in a superposition of states satisfying the constraint of a four-body interaction in the early stage of anneal owing to the optimized schedule. Whether or not such behavior would generically lead to better performance in other problems is an interesting problem.

\section{Discussion}
\label{sec:discussion}
We have adapted the variational method of Refs.~\cite{matsuura2020vanqver,matsuura2020variationally} to the determination of an optimal annealing schedule of the constraint term in the LHZ scheme for quantum annealing in the diabatic regime. Numerical studies of small-size systems with random (spin-glass) and nonrandom (ferromagnetic) interactions reveal that nonmonotonic schedules with a single local minimum lead to significant improvements in the ground-state fidelity in almost all cases. 

It should be noted that the present method is not guaranteed to converge to the global optimum as is usually the case in any variational optimization approach. It is nevertheless encouraging that the fidelity quickly converges to 1 in the ferromagnetic model in Fig.~\ref{fig2b} and improvements in fidelity after ten steps are remarkable as seen in Fig.~\ref{fig6a}.

It is also useful to notice that the study of random problems in Ref.~\cite{brandao2018fixed} based on the quantum approximate optimization algorithm (QAOA) shows that the optimized control parameters for a single instance work fairly well in other instances. Figure \ref{fig6d} indicates that the same may apply to the present case. If this is true, we may perform the process of variational optimization just once for a single example (learning stage) and apply the obtained schedule to other instances in the same problem class (generalization stage) with satisfactory results, thus significantly saving time for repeated optimization.

Although the present study was performed numerically on the classical computer due to the lack of quantum hardware for the LHZ scheme, the result may be suggestive for the design of future hardware under development. A similar approach should be applicable to other terms of the Hamiltonian, and it may serve as a generic framework to find the best possible diabatic annealing schedule not just for the LHZ scheme but also for the standard annealing model.
In particular, the present method may greatly mitigate the difficulty of an appropriate time-dependent control of the coefficient of the constraint term, which almost always shows up when we embed a practical problem on the hardware with nearest-neighbor interactions only.

Another advantage is in the short annealing time for a single iteration, which allows the quantum device to operate with less effects of noise and thus with more reliable outputs, as is always the case for variational classical-quantum hybrid algorithms. Although quantum annealing is relatively stable against noise \cite{Albash2015}, it should nevertheless be better to run a quantum processor under an environment closer to isolation.

\begin{acknowledgments}
This paper is based on results obtained from a project commissioned by the New Energy and Industrial Technology Development Organization (NEDO). In some calculations, we have used the QuTip python library \cite{johansson2012qutip,johansson2013qutip}.
\end{acknowledgments}

\end{document}